\documentclass[%
 reprint,
 amsmath,amssymb,
 aps,
]{revtex4-2}

\usepackage{graphicx}
\usepackage{dcolumn}
\usepackage{bm}

\usepackage{soul}

\usepackage{graphicx}
\usepackage{dcolumn}
\usepackage{bm}
\usepackage{svg}
\usepackage{soul}
\usepackage{hyperref}
\hypersetup{
    colorlinks=true,
    linkcolor=blue,
    filecolor=magenta,      
    urlcolor=cyan,
    pdftitle={Overleaf Example},
    pdfpagemode=FullScreen,
    }
\begin{document}


\title{Estimating and decoding coherent errors
of QEC experiments with detector error models}

\author{Evangelia Takou}
\email{evangelia.takou@duke.edu}
\affiliation{Duke Quantum Center, Duke University, Durham, NC 27701, USA}
\affiliation{Department of Electrical and Computer Engineering, Duke University, Durham, NC 27708, USA}

\author{Kenneth R. Brown}
\email{kenneth.r.brown@duke.edu}
\affiliation{Duke Quantum Center, Duke University, Durham, NC 27701, USA}
\affiliation{Department of Physics, Duke University, Durham, NC 27708, USA}
\affiliation{Department of Electrical and Computer Engineering, Duke University, Durham, NC 27708, USA}
\affiliation{Department of Chemistry, Duke University, Durham, NC 27708, USA}

\date{\today}

\begin{abstract}
 Decoders of quantum error correction (QEC) experiments make decisions based on detected errors and the expected rates of error events, which together comprise a detector error model. Here we show that the syndrome history of QEC experiments is sufficient to detect and estimate coherent errors, removing the need for prior device benchmarking experiments.
Importantly, our method shows that experimentally determined detector error models work equally well for both stochastic and coherent noise regimes. We model fully-coherent or fully-stochastic noise for repetition and surface codes and for various phenomenological and circuit-level noise scenarios, by employing Majorana and Monte Carlo simulators. We capture the interference of coherent errors, which appears as enhanced or suppressed physical error rates compared to the stochastic case, and also observe hyperedges that do not appear in the corresponding Pauli-twirled models. Finally, we decode the detector error models undergoing coherent noise and find different thresholds compared to detector error models built based on the stochastic noise assumption. 
\end{abstract}

\maketitle

\section{Introduction}
A promising route towards realizing large-scale fault-tolerant quantum architectures is by utilizing quantum error correction (QEC). QEC encodes logical information redundantly on several physical qubits, and by carefully correcting physical errors, it can suppress the error rate on the logical level. For incoherent errors, which are modeled as stochastic Pauli errors, the logical error rate decays exponentially with the code distance, as long as the physical errors are below threshold~\cite{aharonov1996arxiv,PreskillJMathPhys2002}. Similarly, for coherent errors it has been shown that a threshold also exists~\cite{PreskilNewJPhys2020,Bravyi2018npjQI,MartonQuantum2023,PatoPRA2025}. Coherent errors on the physical level are often more complicated to understand than stochastic errors, since they introduce coherent errors on the logical level~\cite{DuttonQuantumSciTechnol2017,Bravyi2018npjQI}, and they lead to different failure distributions compared to stochastic models~\cite{LucarelliPRA2017}.

Coherent noise which can arise from miscalibration of gates~\cite{BrownPRL2018}, spectator qubits~\cite{WallraffPRApplied2020}, measurement~\cite{UstinovPRApplied2025} or state preparation errors, is of concern for QEC codes as it is typically more detrimental to the performance of QEC codes than stochastic errors~\cite{LaflammePRL2018,SuzukiPRL2017}. Further, the Pauli-twirled coherent channel underestimates the impact of coherent noise on the logical error rate~\cite{BrownPRA2016,Bravyi2018npjQI}. The impact of coherent noise on QEC codes has only been studied in specific setups as it is generally not easy to simulate it efficiently on large QEC codes. In particular, Bravyi \textit{et al}.~\cite{Bravyi2018npjQI}, showed that coherent noise along a single-axis can be efficiently simulated on surface codes in a code-capacity setup, using Majorana simulators. This formalism has been applied to repetition codes~\cite{SuzukiPRL2017},  surface codes and other topological planar codes~\cite{MartonPRA2024,PatoPRA2025,VennPRL2023,VennPRR2020,Beri20204arXiv}, and extended to the case where measurements experience classical readout errors~\cite{MartonQuantum2023}. Recently, Ref.~\cite{ProctorArxiv2025} developed efficient simulation methods to simulate approximately small Markovian errors, including coherent errors, to study the impact of more complex noise models for near-term devices.

These works have opened interesting avenues for further exploration of coherent noise effects on QEC experiments. For example, it is still unclear what type of decoding graphs or hypergraphs coherent noise generates, and how this structure compares to stochastic detector error models (DEMs) that can be generated efficiently by Stim~\cite{GidneyQuantum2021}. In terms of decoding performance, previous works have mainly assumed uniform weights to decode coherent noise via MWPM~\cite{HiggottArxiv2021,HiggottQuantum2025}. This may not be the optimal decoding choice for DEMs undergoing coherent noise, since interference of coherent errors may occur, and further, circuit-level DEMs might also exhibit hyperedges. Finally, although arbitrary coherent errors cannot be simulated both efficiently and exactly, a more important question is: can these errors be estimated efficiently from the syndrome data from QEC experiments so as to configure a DEM, and how important is the impact of coherent noise on the decoding performance?

In this paper, we attempt to address those questions by estimating coherent noise from syndrome data, which reveals the true structure of the DEM, and allows us to observe interference of coherent errors. We perform code-capacity, phenomenological, and finally circuit-level simulations of coherent noise, for repetition and surface codes. Already from the code-capacity simulation of surface codes, we find coherent enhancement of certain error rates for boundary qubits. Remarkably, such enhancement factors can be estimated from syndrome data, without changing our noise estimation methods which hold for Pauli noise, and we can inform the decoder about such features. We also find that the estimated DEM of a rotated surface code undergoing coherent data qubit errors and readout errors leads to suppressed logical error rate compared to a uniform-weight DEM. For coherent noise on data and ancilla qubits for a repetition code, we find a threshold of $\sim 8\%$, which is smaller compared to the threshold of $10.3\%$ obtained from the corresponding stochastic model. When gates additionally suffer from coherent errors, we find that the repetition code DEM contains hyperedges, making decoding more challenging. Such hyperedges do not appear in the corresponding Pauli-twirled models. We find a threshold of $2.5\%$ if those hyperedges are ignored and a uniform-weight decoder is used. 
Finally, the presence of hyperedges complicates the estimation of edge error-rates, since we have to subtract from the statistics the effect of higher-order correlations. By doing so, we obtain a DEM which leads to {\color{black}{the same threshold}} as the uniform-weight DEM, but achieves a reduced logical error rate.


The paper is structured as follows.  In Sec.~\ref{Sec:Overview}, we give an overview of noise-learning methods from syndrome measurements executed during a QEC experiment. In Sec.~\ref{Sec:Results}, we apply our estimation method to learn coherent noise for code-capacity, phenomenological or circuit-level noise on repetition or surface codes, and study the decoding performance of the estimated DEMs.

\section{Overview of detector error model estimation from syndrome history\label{Sec:Overview}}
To decode errors optimally, a decoder needs to be informed about the error rates of several components failing, as well as time or spatial correlations of errors. This information is usually captured by a DEM~\cite{GidneyQuantum2021,EisertArxiv2024}, which lists the detectors that are triggered by an error mechanism and the corresponding probabilities. A DEM can also be represented as a weighted decoding graph or hypergraph, where the nodes are the detectors and the (hyper-)edges are weighted based on the error rates according to $w=\log((1-p)/p)$~\cite{HiggottQuantum2025}. 

One way to construct the DEM of a QEC experiment, is to first 
characterize the device and obtain experimental error rates of the circuit-level model, i.e., the Pauli-error rates of qubits, gates and measurements and feed such imperfect circuit to a DEM generator such as Stim~\cite{GidneyQuantum2021}. Several characterization methods~\cite{FlammiaACES2022, JiangNatCommun2023,HarperPRXQ2025,HarperArxiv2025,FlammiaArxiv2025} can learn the Pauli-error rates of a circuit. Typically these methods use Pauli-twirling to twirl-away any coherent noise, and obtain an effective Pauli noise model for decoding a non-twirled QEC experiment. As such, the decoder does not exploit the full information about the underlying coherent component of the noise, which leaves open questions to how detrimental such a mismatch is for the decoding performance, or what logical error rates should be expected when coherent noise is present. Further, noise characterization methods such as tomography do not scale favorably with the system size~\cite{JianPanPRL2017}, whereas methods relying on randomized benchmarking require pre-optimization of extra circuits~\cite{HarperPRXQ2025} that need to be run on the same QEC device, besides the QEC experiment of interest.

A more direct approach for estimating noise is to directly infer the error rates of the DEM, from the syndrome data obtained from a QEC experiment~\cite{SpitzAdvQTechn2018,YoungArxiv2025,BrownArxiv2025,WallraffArxiv2025}. This avoids the extra experimental cost of noise characterization experiments, and constructs a structure that is directly usable to the decoder, i.e., a DEM. When the DEM represents a decoding graph, all its error mechanisms correspond to edges, which can be bulk or boundary edges. Boundary edges refer to edges that connect to a single detector. Ref.~\cite{SpitzAdvQTechn2018} has shown that bulk edges can be estimated from the syndrome statistics via the expression:
\begin{equation}\label{Eq:Bulk_edge}
    p_{ij}=\frac{1}{2}-\sqrt{\frac{1}{4}-\frac{\langle v_iv_j\rangle -\langle v_i\rangle \langle v_j\rangle}{1-2(\langle v_i\rangle +\langle v_j\rangle)+4\langle v_iv_j\rangle}},
\end{equation}
whereas the boundary edges via the equation:
\begin{equation}\label{Eq:Boundary_edge}
    p_{ii}=\frac{1}{2}+\frac{\langle v_i\rangle -1/2}{\prod_{j\neq i}(1-2p_{ij})}.
\end{equation}
Here $\langle v_i\rangle$ is the average number of times a detector $i$ fires, and $\langle v_iv_j\rangle$ is the average number of coincidence clicks from two detectors $i$ and $j$, across $N$ experimental shots of a QEC experiment. 

So far, previous works~\cite{SpitzAdvQTechn2018,BrownArxiv2025,WallraffArxiv2025} have used these formulas to create DEMs that undergo Pauli-noise. In what follows, we will show how the same noise estimation method can be used to learn coherent noise, and we will additionally inspect the effects of coherent noise on decoding performance. 


\begin{figure*}[!htbp]
    \centering
    \includegraphics[scale=0.7]{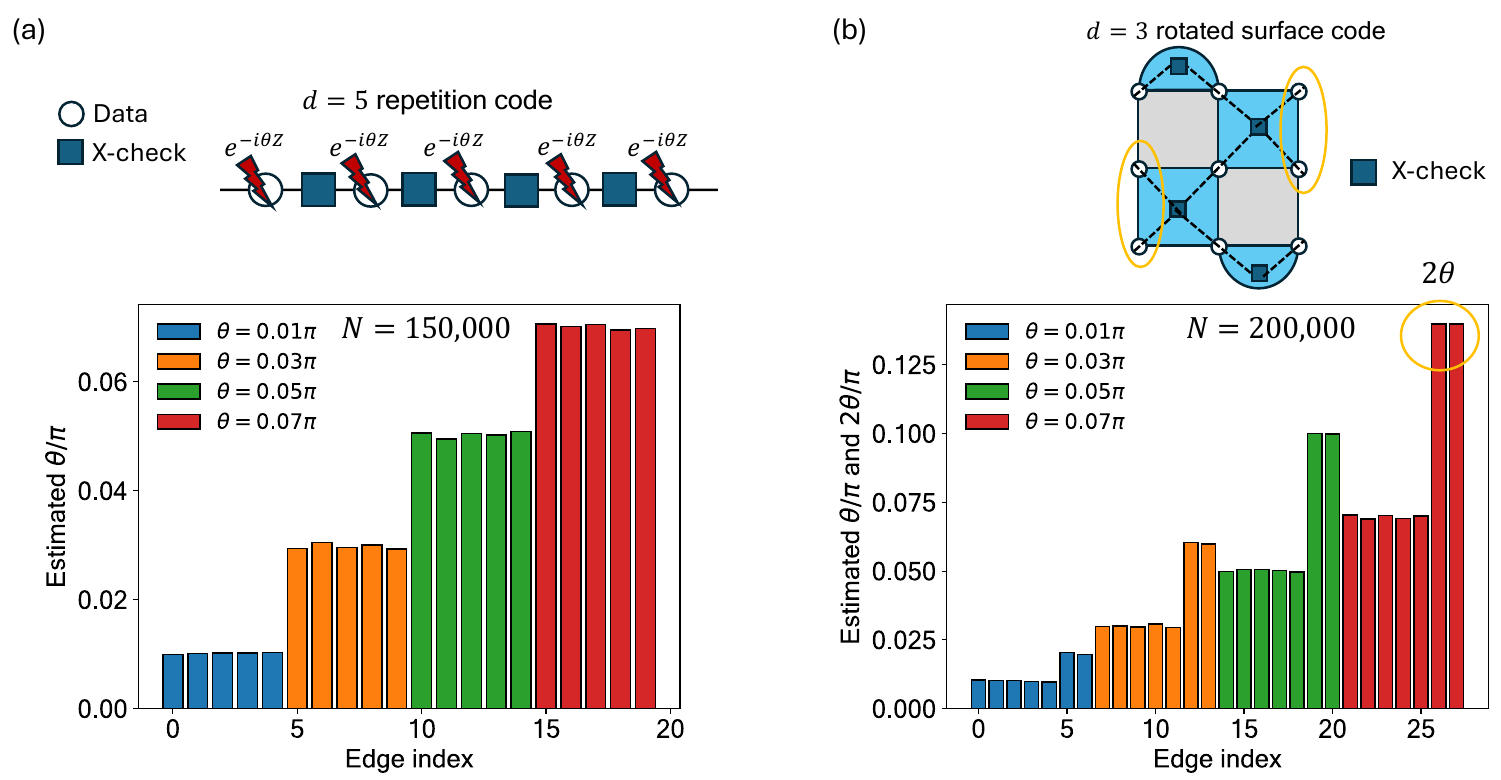}
    \caption{Estimating the error angles $\theta$, when data qubits experience a coherent noise channel $e^{-i\theta Z}$. (a) Estimated angles for $d=5$ $X$-memory repetition code. The bars within each color correspond to different qubits in the same repetition code. The different colors correspond to a different memory simulation where we set a new value for the rotation angles. (b) Estimated angles for the a distance $d=3$ $X$-memory rotated surface code. For boundary edges of the DEM connecting to weight-2 checks, we estimate the cumulative $2\theta$ angle since there is only one detector attached to the two different data qubits, leading to a single boundary edge in the DEM.    }
    \label{fig:Code_cap_est}
\end{figure*}

\section{Estimating coherent noise from syndromes \label{Sec:Results}}

\subsection{Code capacity noise on repetition and surface codes }

We begin the analysis of coherent noise considering a simple setup where only data qubits experience the noise channel {\color{black}{$U=e^{-i\theta Z}=R_z(2\theta)$. Note, that the rotation angle that the qubits experiences on the Bloch sphere is $2\theta$. The Pauli-twirled version of the coherent channel acting on the data qubits is given by:
\begin{equation}
    \mathcal{E}(\rho) = \cos^2(\theta)\rho + \sin^2(\theta) Z\rho Z,
\end{equation}
with the physical error rate being $p=\sin^2\theta$.}} Such single-axis coherent noise can be efficiently simulated with a Majorana simulator~\cite{Bravyi2018npjQI}. To generate the results that we will present shortly, we used the Majorana simulator developed in Ref.~\cite{PatoPRA2025}. We consider first an $X$-memory repetition code. Ancilla qubits, measurements and gates are all assumed to be noiseless (i.e., we assume a code-capacity setup). Our goal is to estimate the error rates of the DEM, which for this model consists only of data qubit edges, as shown in the schematics of Fig.~\ref{fig:Code_cap_est}. By repeating the simulation of the memory experiment for a number of shots $N$, we collect the averages $\langle v_i\rangle$ and $\langle v_iv_j\rangle$ of detector counts, which allows us to estimate the probabilities of bulk edges, $p_{ij}$ [Eq.~(\ref{Eq:Bulk_edge})], and of boundary edges, $p_{ii}$ [Eq.~(\ref{Eq:Boundary_edge})]. Then, using the formula $\theta = \frac{1}{\pi}\sin^{-1}(\sqrt{p})$, we can extract the angle $\theta$.

In Fig.~\ref{fig:Code_cap_est}(a) we consider a $d=5$ $X$-memory repetition code, and show the angle we estimate across the 5 qubits, for  $\theta\in[0.01\pi,0.05\pi,0.1\pi,0.15\pi,0.2\pi]$. Each color corresponds to a different simulation where all data qubits exhibit a $Z$-rotation by the same angle. The number of shots we take to estimate $\theta$ is $N=150,000$. In all cases, we find that the estimated values agree very well with the true rotation angles that we set. Further, for the repetition code under code-capacity coherent noise, we see no coherent enhancement, meaning that the error rates are always $p=\sin^2\theta$ across each DEM edge. This is expected, since there is an exact correspondence between data qubits and space-like edges in the DEM.

A more interesting feature appears when we consider an $X$-memory rotated surface code, where all data qubits experience $e^{-i\theta Z}$ errors [see Fig.~\ref{fig:Code_cap_est}(b)]. Performing again our estimation, we find that for almost all edges we estimate  the expected error rate is approximately $p=\sin^2\theta$ and the expect error angle is very close to $\theta$. However, for boundary edges that connect to a weight-4 check we estimate twice the rotation angle, i.e., an angle of $2\theta$ . This is because the two data qubits contribute both to a single boundary edge in the decoding graph, yielding a total error rate of $p_{\text{coh.}}=\sin^2(2\theta)$. To the contrary, if we assume a Pauli-twirled stochastic model, then such boundaries would display an error rate of $p_{\text{stoch.}}=2\sin^2\theta(1-\sin^2\theta)$, since the two errors combine independently via the formula $p_1+p_2-2p_1p_2$, with $p_1=p_2=\sin^2\theta$. For small angles $\theta$, the lowest-order approximation gives $p_{\text{stoch.}}\approx 2\theta^2$, and $p_{\text{coh.}}\approx 4\theta^2$, meaning that $p_{\text{coh.}}\approx 2p_{\text{stoch.}}$. Notably, we see that even in this simple code-capacity model, we find distinct differences between coherent and stochastic noise models, which are captured by our noise estimation method using only the syndrome information. 

\subsection{Phenomenological noise on surface codes}

\begin{figure}[!htbp]
    \centering
    \includegraphics[scale=0.83]{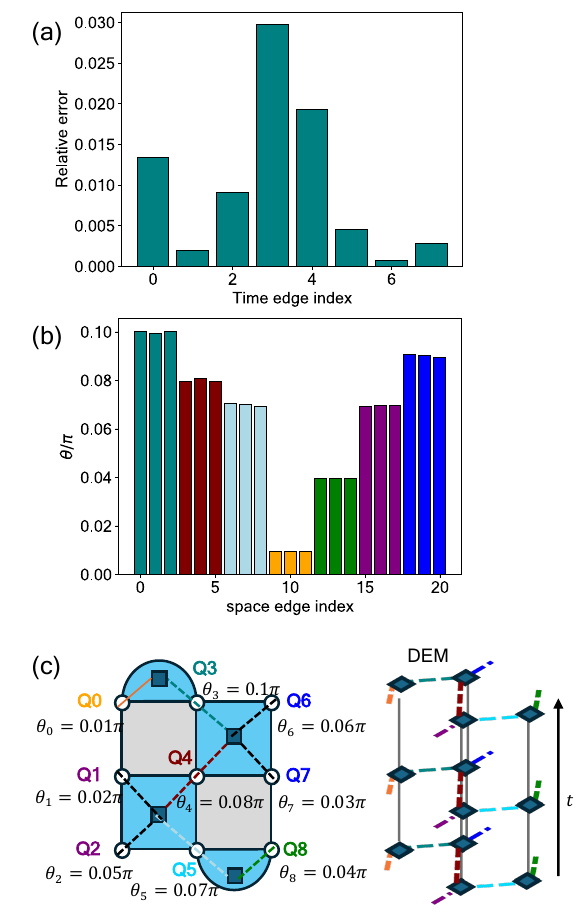}
    \caption{Estimating coherent noise parameters and classical readout errors for a $d=3$ $X$-memory rotated surface code and $r=3$ QEC rounds. (a) Relative error in estimating the error rates of time edges, for a uniform readout error rate of $q=0.03$. (b)  Estimated angles extracted from the estimated error rates of space-like DEM edges. The bars within each color correspond to a different round index. Each color corresponds to a particular edge in the DEM as color-coded in (c). (c) Coherent noise parameters for the data qubits and DEM structure. To estimate the error rates we used $N=180,000$ shots, and for each shot we corrupted the measurement outcome with probability $q=0.03$ for 100 random realizations.   }
    \label{fig:Surface_code_Phenom_est}
\end{figure}

We now consider a phenomenological noise model, where data qubits experience the error channel $U_j=e^{-i\theta_j Z}$ (where $j$ labels the data qubit) per QEC round, while measurements fail classically with some probability $q$ recording the incorrect outcome. Such noise model can still be simulated efficiently with a Majorana simulator~\cite{MartonQuantum2023}. Employing a similar procedure as in Ref.~\cite{MartonQuantum2023}, we simulated a distance $d=3$ $X$-memory rotated surface code  using $N=180,000$ shots, and for each one of them we corrupted the measurement outcomes for  100 random realizations. We set a readout error of $q=0.03$ on all measurements, and spatially varying  angles with $\theta_j\in[0.01\pi,0.1\pi]$, as shown in Fig.~\ref{fig:Surface_code_Phenom_est}(c).

In Fig.~\ref{fig:Surface_code_Phenom_est}(a), we show the relative error $|p_{ij}^{\text{est}.}-p_{ij}^{\text{true}}|/p_{ij}^{\text{true}}$ of estimating the time-like edges of the DEM. For $d=3$ rotated surface code, we have four $X$-checks, and since measurements are repeated for $r=3$ QEC rounds, this gives rise to a total of eight time edges. We find a relative error of less than $\sim 3\%$, ensuring accurate estimation of readout errors. This relative error can be further reduced by increasing the number of shots. In Fig.~\ref{fig:Surface_code_Phenom_est}(b), we show the estimated angles across the different DEM edges. Each color corresponds to a different edge in the DEM according to Fig.~\ref{fig:Surface_code_Phenom_est}(c), and there are three bars per color since we simulate the memory experiment for $r=3$ QEC rounds. For the bulk qubits Q$_3$, Q$_4$, Q$_5$, as well as the boundary qubits Q$_0$, and Q$_8$ that connect to weight-2 checks we recover the $\theta_j$ angle that they experience. For the boundary qubits Q$_1$ and Q$_2$, we find as expected that the sum of the two rotation angles $\theta_1+\theta_2=0.07\pi$, shown with purple bars. Similarly, for the qubits Q$_6$ and Q$_7$, we estimate again $\theta_6+\theta_7=0.09\pi$, as shown with blue bars. Once again, we note the difference between coherent and stochastic models on such boundary qubits of the rotated surface code. In the case of a stochastic noise model, the error rate we would find for a boundary DEM edge connecting to a weight-4 check would be $p_1+p_2-2p_1p_2$, since such errors would combine independently.

\begin{figure}[!htbp]
    \centering
    \includegraphics[scale=0.6]{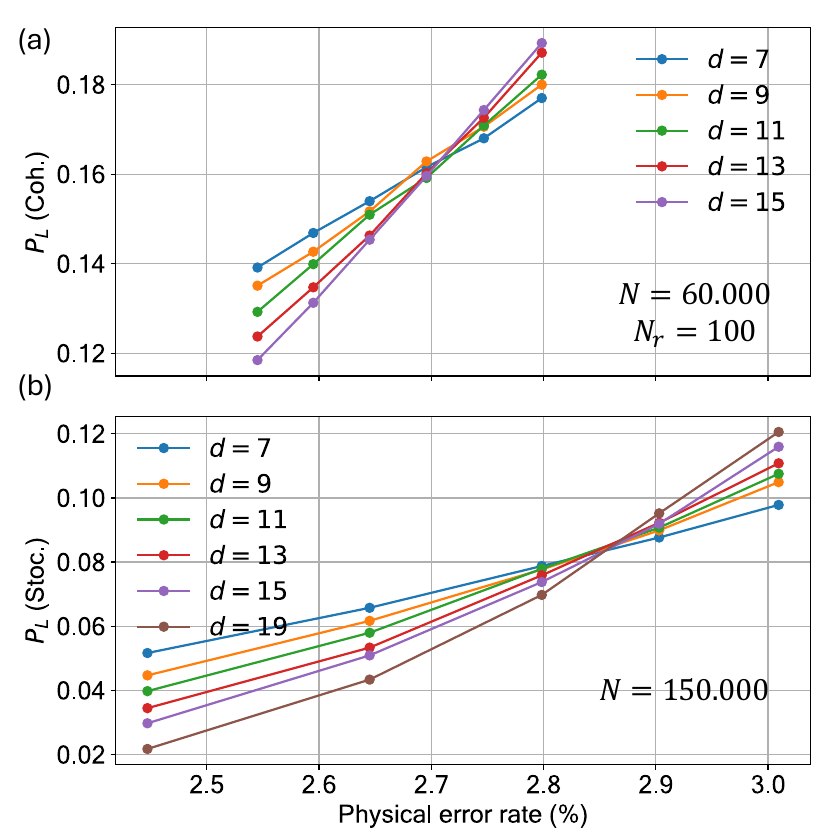}
    \caption{Comparing the logical error rate performance under stochastic or coherent noise models for a rotated surface code, when errors are decoded with uniform-weight decoder. (a) Logical error rate as a function of the physical error rate $p=\sin^2\theta$ when data qubits experience $e^{-i\theta Z}$ and measurements experience readout errors with $q=p$. The number of shots used for the decoding is $N=60,000$, and
    for each one of this shots, we corrupt the measurement outcomes for $100$ different realizations.
    (b) Logical error rate when data qubits experience stochastic $Z$ errors, and measurements experience readout errors. The number of shots used for the decoding is $N=150,000$. }
    \label{fig:TwirledVSCoh_SC}
\end{figure}

To explore how these differences in the physical error rates impact the logical error rate performance, we consider an $X$-memory rotated surface code and the same phenomenological noise model as before,  and compare it to its corresponding Pauli-twirled model. For simplicity, we assume all physical rotation angles to be equal. For the stochastic noise model, we use Stim~\cite{GidneyQuantum2021}, and set only $Z$-errors per round on the data qubits, as well as classical readout errors. We also use $N=150,000$ shots and set a uniform decoding graph (although Stim's DEM sets the boundary probabilities of boundary edges connecting to weight-4 checks as $p_1+p_2-2p_1p_2$). The reason why we assume a uniform-weight decoding graph is because we want to use the same graph for the coherent and stochastic noise models so as to compare their logical  performance faithfully. To decode the DEMs, we  use Pymatching~\cite{HiggottArxiv2021,HiggottQuantum2025}. For the coherent noise model we set the number of shots to $N=60,000$, and corrupt the measurements for 100 random realizations per shot, with uniform probability $q=p=\sin^2\theta$. The logical error rate, $P_L$, for this model corresponds to the infidelity, calculated using the logical rotation angle that the final state experiences after correction~\cite{MartonQuantum2023}. The two logical error rates are compared in Fig.~\ref{fig:TwirledVSCoh_SC}, where we find a threshold $\sim 2.85\%$ for the stochastic model, and $\sim 2.7\%$ for the coherent noise model. 
Although the threshold for the coherent noise model has been reported previously in Ref.~\cite{MartonQuantum2023}, here, we explain why it is smaller compared to the threshold obtained by the stochastic error model. This feature is owned to the fact that certain boundary qubits experience a higher error rate in the coherent noise scenario due to interference of coherent errors.

An interesting feature appears if instead of using uniform decoding graphs, we use the estimated DEMs to decode coherent noise. As we mentioned previously, boundary edges connecting to weight-4 checks experience a cumulative coherent angle error of $2\theta$. To exploit this information for the decoding, we now estimate all edge probabilities to reconstruct the DEM and then calculate the logical error rate (infidelity) $P_L$. We use $N=200,000$ shots for the estimation, where for each shot we corrupt for 100 different realizations the measurement outcomes with error rate $q=p=\sin^2\theta$. Then, we sample a new batch of $N=40,000$ shots (and again corrupting the measurements for $100$ realizations), to collect the detection outcomes and decode using our estimated DEM. 

\begin{figure}[!htbp]
    \centering
    \includegraphics[scale=0.6]{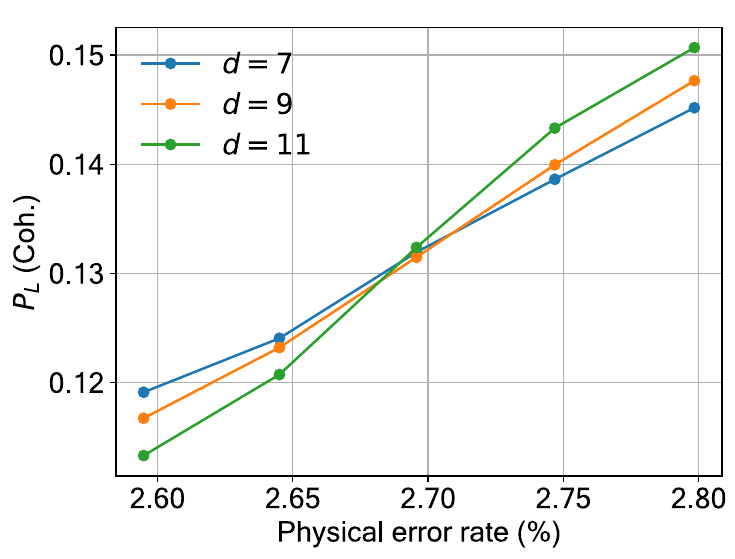}
    \caption{Logical error rate as a function of the physical error rate, when data qubits experience $e^{-i\theta Z}$ errors per QEC round, and measurement outcomes are incorrectly recorded to the opposite bit value with probability $q=p=\sin^2\theta$, for an $X$-memory rotated surface code. Our estimated DEM is constructed from $N=200,000$ syndrome data, where for each shot the outcomes are corrupted with probability $q$ for $100$ random realizations. For the decoding, we use a new batch of $N=40,000$ shots, with detection outcomes corrupted again for another $100$ random realizations per shot. }
    \label{fig:Phenom_Est_DEM}
\end{figure}

In Fig.~\ref{fig:Phenom_Est_DEM}, we show the infidelity, $P_L$, as a function of the physical error rate $p=\sin^2\theta$, where the readout error probability is set to $q=p$. We find again the same threshold at $\sim 2.7\%$, but now $P_L$ has reduced compared to Fig.~\ref{fig:TwirledVSCoh_SC}(a). For example, for a physical error rate at $\sim 2.6\%$, decoding with uniform weights gives a logical error rate at $0.14$ for $d=11$, and slightly higher for $d=7$ and $d=9$. On the other hand, using our estimated DEM, we find that $P_L$ for a physical error rate of $\sim 2.6\%$ is lower than $0.12$, for $d\in[7,9,11]$. Thus, 
we see that by exploiting the information provided to us by the syndrome data, we can capture coherent effects to inform the decoders and improve the logical performance.

\subsection{Circuit-level noise on repetition codes}

\begin{figure}[!htbp]
    \centering
    \includegraphics[scale=0.75]{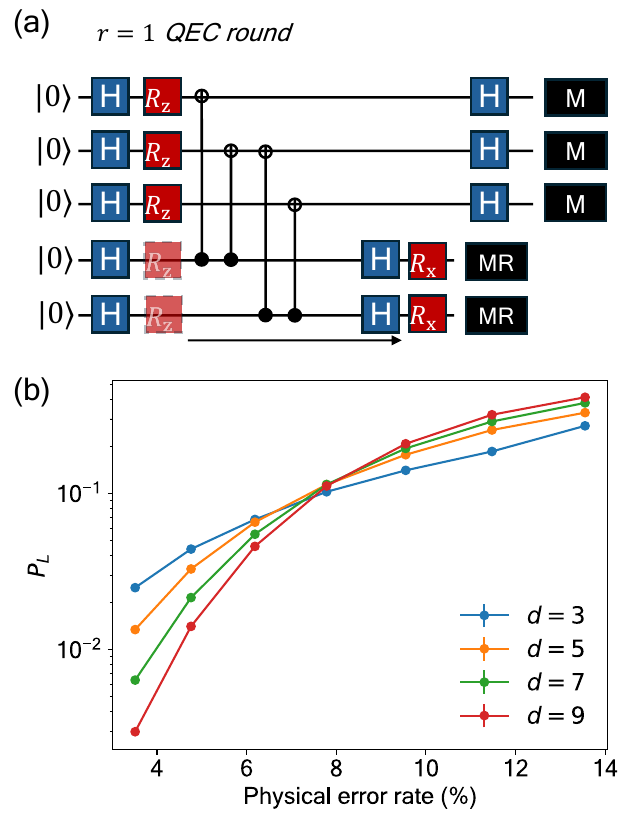}
    \caption{(a) 
    Distance $d=3$ $X$-memory repetition code circuit, where both data and ancilla qubits experience $e^{-i\theta Z}=R_z(2\theta)$ errors at the start of every  QEC round. Here we show $r=1$ QEC round. (b) Logical error rate as a function of the physical error rate $p=\sin^2\theta$, for the error model shown in (a) and $r=d$ QEC rounds. The threshold is slightly smaller than $8\%$. The decoding graph is formed based on the estimated error rates. The number of shots used for estimation and decoding is $N=10^6$. }
    \label{fig:Rep_code_coher_phenom}
\end{figure}

We now turn our attention to circuit-level simulations of coherent noise. Unfortunately generic coherent noise channels cannot be simulated both exactly and efficiently, such as coherent errors that occur on gates or ancillas.   To simulate circuit-level coherent errors, we developed a C++ Monte Carlo simulator. We start with an $X$-memory repetition code, and a noise model where data and ancilla qubits experience $e^{-i\theta Z}$ errors per QEC round. The errors on ancilla qubits could be attributed to coherent initialization errors, or if propagated to the measurements as shown in Fig.~\ref{fig:Rep_code_coher_phenom}(a), can be thought of as coherent measurement errors. In the latter case, these measurement errors imply that we are no longer  measuring the ideal $X_iX_j$ stabilizers.

To calculate the logical error rate, we measure the data qubits in the final round, and then check if the inferred correction obtained from MWPM and the actual errors both flip (or do not flip) the logical operator $X_L$. The reason why we follow this approach and do not use the maximum infidelity (or diamond norm to the identity), is because the state can leak outside of the codespace, even after correction.  Figure~\ref{fig:Rep_code_coher_phenom}(b) shows the logical error rate as a function of the physical error rate (assuming all $\theta_j$ are equal for all qubits), for distances $d\in[3,5,7,9]$. We decode the DEM using the estimated weights (the same performance is obtained if we use a uniform graph). We find a threshold of $\sim 8\%$, which is lower compared to the stochastic noise model which leads to a threshold of $10.3\%$. This difference in thresholds is attributed to coherent measurement errors, and how these combine with data qubit errors, since as we mentioned in the pure code-capacity setup the coherent and Pauli-twirled models for a repetition code memory have the same thresholds. 


\begin{figure}[!htbp]
    \centering
    \includegraphics[scale=0.7]{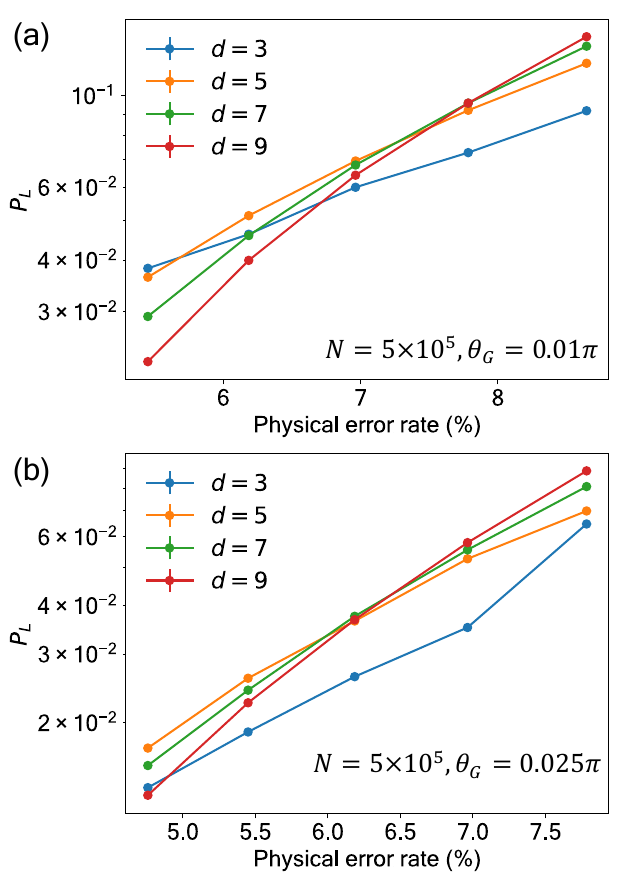}
    \caption{(a) Logical error rate as a function of the physical error rate, for an $X$-memory repetition code, where qubits experience $e^{-i\theta Z}$ errors per QEC round, and perfect CNOT gates are followed by $e^{i\theta_G Z_cZ_t}$ error. The gate error is fixed to $\theta_G=0.01\pi$, and the number of shots used to estimate the DEM and decode is $N=5\times 10^5$. (b) Same as in (a) for a gate error angle $\theta_G=0.025\pi$.}
    \label{fig:LER_Fixed_gate_error}
\end{figure}

\begin{figure*}[!htbp]
    \centering
    \includegraphics[scale=0.75]{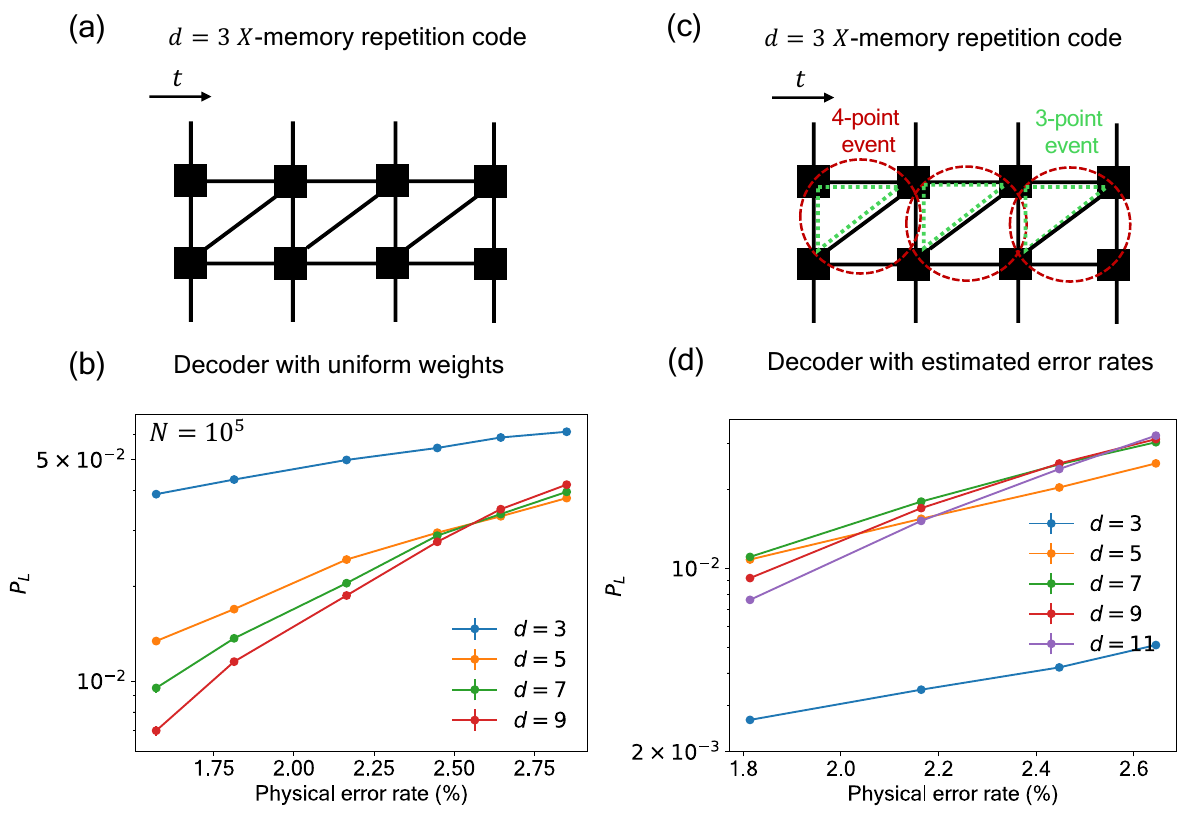}
    \caption{(a) Detector error model of $d=3$ $X$-memory repetition code, for coherent qubit errors $e^{-i\theta Z}$, and coherent gate errors $e^{i\theta_G Z_cZ_t}$, assuming no higher-order events. (b) Logical error rate as a function of the physical error rate for $\theta_{\text{data}}=\theta_{\text{anc}}=\theta_G$, and $N=10^5$ decoding shots, assuming a uniform decoding graph as in (a). (c) Detector error model of $d=3$ $X$-memory repetition code, with three-, and four-order events. (d) Logical error rate as a function of the physical error rate, for a DEM constructed based on estimated error rates. Higher-order events are used to only redefine the error rates of bulk and boundary edges, and are then ignored. {\color{black}{The number of shots for the estimation and decoding is $N=10^5$ for $d=11$, and $N=10^6$ for $d\in[3,5,7,9]$.}} }
    \label{fig:Rep_code_w_gate_errors}
\end{figure*}

As a next example, we consider a more complete noise model, where gates also experience coherent errors. We simulate such gate errors by assuming perfect CNOT gates followed by $e^{i\theta_G Z_cZ_t}$ errors, where $c$ denotes the control qubit, $t$ the target qubit, and $\theta_G$ is the error angle. Because of the coherent gate errors, we find that higher-order detection events (meaning hyperedges)  exhibit non-zero error rates. If $\theta_G$ is small compared to $\theta_{\text{data}}$ and $\theta_{\text{anc}}$, then one can ignore the higher-order correlations and decode the DEM based on the estimated error rates of edges. In Fig.~\ref{fig:LER_Fixed_gate_error}(a) we show the logical error rate for $\theta_G=0.01\pi$ and in Fig.~\ref{fig:LER_Fixed_gate_error}(b) for $\theta_G=0.025\pi$. 
The $d=3$ line exhibits finite-sized effects. For $\theta_G=0.01\pi$, the threshold is close to $7.5\%$ whereas for $\theta_G=0.025\pi$ the threshold is close to $6.25\%$.
As expected, for a small rotation angle $\theta_G$ and upon truncating our estimation method to  two-point probabilities (i.e., edges), the threshold reduces as $\theta_G$ increases. However, as $\theta_G$ increases further and becomes comparable with $\theta_{\text{data}}$ and $\theta_{\text{anc}}$, we will eventually obtain incorrect probabilities for the edges (and even negative values) due to the presence of hyperedges. For this reason, we need to estimate the error rates of hyperedges and use them to redefine the lower-order probabilities. We should further mention that although Ref.~\cite{SuzukiPRL2017} studied a similar circuit-level noise model, the authors did not study the structure of the repetition code DEM, and hence, did not identify the presence of hyperedges.

To include the effect of higher-order events, we follow a similar approach as in Ref.~\cite{BrownArxiv2025}. Specifically, we calculate three-, and four-point probabilities as shown in Fig.~\ref{fig:Rep_code_w_gate_errors}(c), and use those values to redefine the probabilities of bulk and boundary edges, by recursively applying the formula:
\begin{equation}
    p_{\text{lower-order}}' = \frac{p_{\text{lower-order}}-p_{\text{higher-order}}}{1-2 p_{\text{higher-order}}}.
\end{equation}
Given a subset of detectors $(D_i,D_j)$ that form a bulk edge, or a detector $D_k$ associated with a boundary edge, we inspect if such detector set is a subset of the higher-order event. If this happens, then we remove the contribution of the higher-order event from the lower-order probability. Note that we have not exhaustively checked all higher-order events that span the entire DEM structure. Further, we only subtract contributions from one three-point event and one four-point event per edge probability. This choice is not unique, but we simply found that this choice leads to a DEM with better logical performance  than the uniform weight DEM, as we will shortly show. After redefining lower-order probabilities, we choose to ignore the higher-order events (i.e., we do not perform a decomposition to include them in the DEM), and then, we decode the estimated DEM with MWPM. 

In Fig.~\ref{fig:Rep_code_w_gate_errors}(b), we show the logical error rate as a function of the physical error rate for $\theta_{\text{data}}=\theta_{\text{anc}}=\theta_G$ and $N=10^5$ shots, when a uniform DEM is used, and the hyperedges are ignored. In this case, we find a threshold close to $2.5\%$. We also note that the $d=3$ case exhibits finite-size effects. In Fig.~\ref{fig:Rep_code_w_gate_errors}(d), we show again the logical error rate, when we use the estimated DEM, for which the number of shots for the estimation and decoding is set to {\color{black}{$N=10^5$ for $d=11$, and $N=10^6$ for the remaining distances. In this case, we observe a threshold close to $2.5\%$, as for the uniform-weight decoder, but we additionally find a reduction in the logical error rate. }}

Overall, through this repetition code simulation, we find that coherent noise can lead to entirely different DEM structures compared to stochastic Pauli noise. Such DEM structure is completely missed when noise characterization is performed on Pauli- or Clifford-twirled benchmarking circuits, and then used for decoding a non-twirled QEC experiment. By directly using the syndrome information of a QEC experiment one could potentially alleviate such mismatch of detector error models. Testing 
general QEC codes under coherent noise and inspecting further the decompositions of hyperedges that could appear in their DEMs [or using hyper-graph decoders such as as Ref.~\cite{ShrutiArXiv2025}] is an interesting future avenue for gauging the impact of coherent errors in the logical performance.

\section{Conclusions}
Coherent noise can be more detrimental to the logical performance of QEC experiments than Pauli noise. Learning and analyzing coherent errors that unavoidably appear in QEC experiments is important for extracting informed error models and configuring decoders to this type of noise. In this work, we studied repetition and surface codes and  showed that coherent noise can lead to interference effects or even entirely different structures of detector error models compared to their stochastic noise counterparts. Importantly, such features, can be observed  from the syndrome history alone. For a rotated surface code under single-axis coherent noise and readout errors, we showed that boundary DEM edges experience a higher physical error rate, and that the logical error rate is reduced if we use the estimated DEM instead of a uniform-weight DEM. For a repetition code under coherent data and ancilla errors we showed that the threshold reduces by $\sim 2\%$ compared to the stochastic noise model. When coherent gate errors are also present, the estimated DEM contains hyperedges, which is a feature not observed in the Pauli-twirled model. {\color{black}{Using our estimated DEM, we were able to reduce the logical error rate, while preserving the same threshold as the uniform-weight decoder.}} Through these examples we showed how our noise estimation method can reconstruct DEMs that undergo coherent or stochastic noise, without incurring extra experimental overhead. 
Our methods are simple and efficient and can be directly applied to QEC experiments to learn noise and improve the logical error suppression.

\acknowledgements{
The authors would like to thank Balint Pato for useful discussions. This work was supported by the Office of the Director of National Intelligence (ODNI), Intelligence Advanced Research Projects Activity (IARPA), under the Entangled Logical Qubits program through Cooperative Agreement Number W911NF-23-2-0216 and the ARO/LPS QCISS program (W911NF-21-1-0005).
}




\bibliography{apssamp}

\end{document}